%% file: BAD-1619.tex
\documentclass[twocolumn,showpacs,aps,prl,superscriptaddress]{revtex4}

\usepackage[dvips]{graphicx}
\usepackage{dcolumn}
\usepackage{epsfig}
\usepackage{amsmath}
\RequirePackage{xspace}

\usepackage{relsize}
\def\babar{\mbox{\slshape B\kern-0.1em{\smaller A}\kern-0.1em
    B\kern-0.1em{\smaller A\kern-0.2em R}}}
\def\CP                {\ensuremath{C\!P}\xspace}
\def\ra                 {\ensuremath{\rightarrow}\xspace}
\def\to                 {\ensuremath{\rightarrow}\xspace}

\def\Bbar    {\kern 0.18em\overline{\kern -0.18em B}{}\xspace}

\def\BB      {\ensuremath{B\Bbar}\xspace} 
\def\epem       {\ensuremath{e^+e^-}\xspace}

\newcommand{\mev}{\ensuremath{\mathrm{\,Me\kern -0.1em V}}\xspace}

\def\Bz      {\ensuremath{B^0}\xspace}

\newcommand{\etapr}{\ensuremath{\eta^{\prime}}\xspace}
\newcommand{\gev}{\ensuremath{\mathrm{\,Ge\kern -0.1em V}}\xspace}
\def\Bzb     {\ensuremath{\Bbar^0}\xspace}

\newcommand{\msp}{\ensuremath{\phantom{-}}}

\newcommand{\fetapreppggkzl}{\ensuremath{\etapr_{\eta(\gamma\gamma)\pi\pi} \KL }}

\newcommand{\fetapreppggk}{\ensuremath{\etapr_{\eta(\gamma\gamma)\pi\pi} K^{+} }}

\newcommand{\fetapreppthrpik}{\ensuremath{\etapr_{\eta(3\pi)\pi\pi} K^{+}}}

\newcommand{\etappp}{\ensuremath{\eta_{3\pi}}}
\newcommand{\etaprg}{\ensuremath{\etapr_{\rho\gamma}}}

   \newcommand{\rhoz}{\ensuremath{\rho^0}}

\newcommand{\etapeppgg}{\ensuremath{\etapr_{\eta(\gamma\gamma)\pi\pi}}}
\newcommand{\etapeppppp}{\ensuremath{\etapr_{\eta(3\pi)\pi\pi}}}

\def\pip   {\ensuremath{\pi^+}\xspace}
\def\pim   {\ensuremath{\pi^-}\xspace}

\def\Bz      {\ensuremath{B^0}\xspace}
\def\Bzb     {\ensuremath{\Bbar^0}\xspace}
\def\BzBzb   {\ensuremath{\Bz {\kern -0.16em \Bzb}}\xspace}
\def\Bu      {\ensuremath{B^+}\xspace}
\def\Bub     {\ensuremath{B^-}\xspace}

\def\BpBm    {\ensuremath{\Bu {\kern -0.16em \Bub}}\xspace}

\providecommand{\DE}{\ensuremath{\Delta E}}
\providecommand{\pvec}{{\bf p}}
\providecommand{\half}{\mbox{${1\over2}$}}

\providecommand{\etapKz}{\mbox{$\eta^{\prime} K^0$}}
\providecommand{\etapKzs}{\mbox{$\eta^{\prime} \KS$}}
\providecommand{\etapKzl}{\mbox{$\eta^{\prime} \KL$}}

\providecommand{\skz}{\mbox{$S$}}
\providecommand{\ckz}{\mbox{$C$}}

\providecommand{\UfourS}{\mbox{$\Upsilon(4S)$}}
\providecommand{\BetapKz}{\mbox{$B^0 \rightarrow \eta^{\prime} K^0$}}
\providecommand{\BetapKzs}{\mbox{$B^0 \rightarrow \eta^{\prime} \KS$}}
\providecommand{\BetapKzl}{\mbox{$B^0 \rightarrow \eta^{\prime} \KL$}}
\newcommand{\SetapKz}{\ensuremath{0.58\pm 0.10\pm 0.03}}
\newcommand{\CetapKz}{\ensuremath{-0.16\pm 0.07\pm 0.03}}
\newcommand{\etal}{{\em et al.}}

\def\deltat{\ensuremath{{\rm \Delta}t}\xspace}
\def\deltaS{\ensuremath{{\rm \Delta}S}\xspace}
\providecommand{\dt}{\deltat}
\def\deltamd{\ensuremath{{\rm \Delta}m_d}\xspace}
\def\stwob{\ensuremath{\sin\! 2 \beta   }\xspace}

\def\KS    {\ensuremath{K^0_{\scriptscriptstyle S}}\xspace}
\def\KL    {\ensuremath{K^0_{\scriptscriptstyle L}}\xspace}

\def\gaga  {\ensuremath{\gamma\gamma}\xspace}
\def\mes        {\mbox{$m_{\rm ES}$}\xspace}

\def\piz{\mbox{${\pi^{0}}$}}

\def\qqbar{\mbox{$q\bar q\ $}}

\def\Bub     {\ensuremath{B^-}\xspace}
\def\Bu      {\ensuremath{B^+}\xspace}
\def\B{\mbox{$B$}}
\providecommand{\tcp}{\mbox{$t_{\CP}$}}
\def\BzBzb   {\ensuremath{\Bz {\kern -0.16em \Bzb}}\xspace}
\def\BpBm    {\ensuremath{\Bu {\kern -0.16em \Bub}}\xspace}
\newcommand{\gevc}{\ensuremath{{\mathrm{\,Ge\kern -0.1em V\!/}c}}\xspace}
\newcommand{\mevc}{\ensuremath{{\mathrm{\,Me\kern -0.1em V\!/}c}}\xspace}
\newcommand{\gevcc}{\ensuremath{{\mathrm{\,Ge\kern -0.1em V\!/}c^2}}\xspace}
\newcommand{\mevcc}{\ensuremath{{\mathrm{\,Me\kern -0.1em V\!/}c^2}}\xspace}

\newcommand{\ttag}{\ensuremath{t_{\rm tag}}}

\newcommand{\etagg}{\ensuremath{\eta_{\gaga}}}
\newcommand{\mb}{\mes}

\newcommand{\xf}{\mbox{${\cal F}$}}
\newcommand{\bflav}{\ensuremath{B_{\rm flav}}}
\providecommand{\sigdt}{\ensuremath{\sigma_{\deltat}}}

\newcommand{\fetaprgKp}{\ensuremath{\etapr_{\rho\gamma} K^+}}

\newcommand{\fetaprgKz}{\ensuremath{\etapr_{\rho\gamma} K^0}}

\newcommand{\fetapKz}{\ensuremath{\etapr K^0}}

\newcommand{\thetaT}{\ensuremath{\theta_{\rm T}}}

\def\pep2{PEP-II}

\newcommand{\jprlBase}       {Phys.\ Rev.\ Lett.\xspace}
\newcommand{\jprl}      [1]  {\jprlBase\ {\bf #1}}

\newcommand{\jprBase}        {Phys.\ Rev.\xspace}
\newcommand{\jprd}      [1]  {\jprBase\ D~{\bf #1}}

\newcommand{\progtp}    [1]  {{Prog.\ Theor.\ Phys.\ {\bf #1}}}
\newcommand{\jplBase}        {Phys.\ Lett.\xspace}
\newcommand{\plb}       [1]  {\jplBase\ B~{\bf #1}}
\newcommand{\npBase}         {Nucl.\ Phys.\xspace}
\newcommand{\npb}       [1]  {\npBase\ B~{\bf #1}}
\newcommand{\nimBaseA}       {Nucl.\ Instr.\ Meth.\xspace}
\newcommand{\nima}      [1]  {\nimBaseA~A~{\bf #1}}
\newcommand{\jpg}       [1]  {{J.\ Phys.\ {\bf G{\bf #1}}}}

\def\Kz    {\ensuremath{K^0}\xspace}

\newcommand{\BABARPubYear}    {06}
\newcommand{\BABARPubNumber}  {063}

\newcommand{\SLACPubNumber} {12127}

\begin{document}

\preprint{\babar-PUB-\BABARPubYear/\BABARPubNumber} 
\preprint{SLAC-PUB-\SLACPubNumber} 

\begin{flushleft}
\babar-PUB-\BABARPubYear/\BABARPubNumber\\
SLAC-PUB-\SLACPubNumber\\
%hep-ex/\LANLNumber\\[10mm]
\end{flushleft}

\title{ \large \bf\boldmath  Observation of \CP\ violation in
  \BetapKz\  Decays } 

\input authors_sep2006.tex

\begin{abstract}
We present measurements of the 
time-dependent \CP-violation parameters \skz\ and \ckz\ in $\B^0\ra \etapr \Kz$ decays.
The data sample corresponds to 384 million \BB\ pairs 
produced by \epem\ annihilation at the \UfourS .
The results are $\skz = \SetapKz$, and $\ckz = \CetapKz$. We observe 
mixing-induced  \CP violation with a significance of 5.5 standard deviations in this $b\ra s$
penguin dominated mode.   
\end{abstract}

% PACS, the Physics and Astronomy Classification Scheme.

\maketitle
Measurements of time-dependent \CP\  asymmetries in $B^0$ meson decays through
 Cabibbo-Kobayashi-Maskawa (CKM) favored $b \rightarrow c \bar{c} s$ amplitudes
\cite{babar} have provided  crucial tests of the mechanism of \CP\ violation
in the Standard Model (SM) \cite{SM}.
Decays of $B^0$ mesons to charmless hadronic final states such as
$\etapr K^0$ proceed mostly via a single loop (penguin)
amplitude.  In the SM the penguin amplitude has approximately  the same
weak phase as the $b \to c \bar{c} s$ transition, but it is sensitive to
the possible presence of new heavy particles in the loop
\cite{Penguin}.  The measurement of \CP\  asymmetries in
\BetapKz\ thus provides an important test for such effects.  

Within the SM,
CKM-suppressed  amplitudes and multiple 
particles in the  
loop introduce additional weak phases whose contribution may not be negligible
\cite{Gross,Gronau,BN,london}.
The time-dependent \CP-violation parameter $S$ (defined in
Eq.~\ref{eq:FCPdef} below) measured in the decay \BetapKz\ is compared
with the value of \stwob\ from measurements of time-dependent
\CP\ violation in $B$ decays to states containing charmonium and a
neutral kaon.  The deviation $\deltaS=S - \stwob$
has been estimated in
several theoretical approaches: QCD factorization (QCDF) \cite{BN,BN2}, QCDF with modeled
rescattering \cite{Cheng}, Soft Collinear Effective Theory \cite{Zupan},
and SU(3) symmetry \cite{Gross,Gronau, Jonat}.
 These models estimate $|\deltaS|$ to be of the order $0.01$, and with
uncertainties give bounds  $|\deltaS | \lesssim 0.05$.

The time-dependent \CP\ asymmetry in the decay \BetapKzs\ has
been measured previously by the \babar ~\cite{Previous} and Belle~\cite{BELLE} 
Collaborations. 
In this Letter we update our previous  measurements using an  integrated
luminosity of 349~fb$^{-1}$, corresponding to $384 \pm 4$ million \BB\
pairs, recorded at the $\Upsilon (4S)$ resonance (center-of-mass energy
$\sqrt{s}=10.58\ \gev$). Belle has since updated their results
\cite{NewBelle}.
Our  data were collected with the
\babar\  detector~\cite{BABARNIM} at the PEP-II asymmetric-energy
\epem\ collider.
In addition to the \BetapKzs\ decays used previously, we now also include the
decay \BetapKzl.

Charged particles from \epem\ interactions are detected, and their
momenta measured, by a combination of five layers of double-sided
silicon microstrip detectors and a 40-layer drift chamber,
both operating in the 1.5~T magnetic field of a superconducting
solenoid. Photons and electrons are identified with a CsI(Tl)
electromagnetic calorimeter (EMC). Charged particle
identification  is provided by the average energy loss  in
the tracking devices and by an internally reflecting ring imaging
Cherenkov detector covering the central region.  The instrumented 
flux return (IFR) of the magnet allows the identification of muons and 
\KL\ mesons.

We reconstruct a \Bz decaying into the
\CP\ eigenstate $\etapKzs$ or $\etapKzl$ ($B_{\CP}$).  From the
remaining particles in the event we also reconstruct the decay
vertex of the other $B$ meson ($B_{\rm tag}$) and identify its flavor.
The difference $\deltat \equiv \tcp - \ttag$ of the proper decay times
$\tcp$ and $\ttag$ of the \CP\ and tag $B$ mesons, respectively, is
obtained from the measured distance between the $B_{\CP}$ and $B_{\rm
tag}$ decay vertices and from the boost ($\beta \gamma =0.56$) of the
\epem system.
The \deltat\ distribution is given by:
\begin{eqnarray}
  F(\dt) &=&
        \frac{e^{-\left|\deltat\right|/\tau}}{4\tau} [1 \mp\Delta w \pm
                                                   \label{eq:FCPdef}\\
   &&\hspace{-1em}(1-2w)\left(-\eta S\sin(\deltamd\deltat) -
C\cos(\deltamd\deltat)\right)]\,\nonumber
\end{eqnarray}
where $\eta$ is the \CP\ eigenvalue of the final
state ($-1$ for \etapKzs, $+1$ for \etapKzl).
The upper (lower) sign denotes a decay accompanied by a \Bz (\Bzb) tag,
$\tau$ is the mean $\Bz$ lifetime, $\deltamd$ is the mixing frequency,
and the mistag parameters $w$ and $\Delta w$ are the average and
difference, respectively, of the probabilities that a true $\Bz$\ is
incorrectly tagged as a $\Bzb$\ or vice versa.  The tagging algorithm
has six mutually exclusive tagging categories and a measured analyzing 
power  of $( 30.4\pm 0.3)\%$ \cite{s2b}.  A non-zero value of the
parameter $C$ would indicate direct \CP\ violation.

We establish the event selection criteria with the aid of a detailed
Monte Carlo (MC) simulation of the \B\ production and decay sequences,
and of the detector response \cite{geant}.  
These criteria are designed
to retain signal events with high efficiency while removing most of
the background.  

The \B-daughter candidates are reconstructed through their decays
$\piz\ra\gaga$, $\eta\ra\gaga$ (\etagg), $\eta\ra\pip\pim\piz$
(\etappp), $\etapr\ra\etagg\pip\pim$ (\etapeppgg), 
$\etapr\ra\etappp\pip\pim$ (\etapeppppp),
$\etapr\ra\rhoz\gamma$
(\etaprg), where $\rhoz\ra\pip\pim$, $\KS\ra\pip\pim$ ($K^0_{\pi^+\pi^-}$) or
$\piz\piz$ ($K^0_{\pi^0\pi^0}$).  Only the \etapeppgg\ mode is used for
the \etapKzl\ sample.
The requirements on the invariant masses of these particle combinations
are the same as in our previous analysis \cite{Previous}.
The list of all decay modes used in the current analysis can be seen in
Table \ref{tab:Results}.  
Signal \KL\ candidates are reconstructed from clusters of energy
deposited in the EMC or from hits in the IFR not associated with any
charged track in the event  \cite{Stefan}.  From the cluster centroid and the \Bz
decay vertex we determine the direction (but not the magnitude) of the
$\KL$ momentum $\pvec_{\KL}$.

For \etapKzs\ decays we reconstruct the \B-meson candidate by
combining the four-momenta of the \KS\ and \etapr\ with a
vertex constraint.  
We also constrain the  $\eta$, \etapr, and \piz\ masses to world-average values
\cite{PDG2006}.
From the kinematics of \UfourS\ decays we determine the
energy-substituted mass $\mes \equiv \sqrt{(\half s +
\pvec_0\cdot\pvec_B)^2/E_0^2 - \pvec_B^2}$ and the energy difference
$\DE \equiv E_B^*-\half\sqrt{s}$, where $(E_0,\pvec_0)$ and
$(E_B,\pvec_B)$ are the laboratory four-momenta of the \UfourS\ and the $B$ candidate,
respectively, and the asterisk denotes the \UfourS\ rest frame.  The
resolution is $3\ \mev$ in \mes\ and $20-50\ \mev$ in \DE,
depending on the decay mode. 

For \etapKzl\ candidates we obtain \DE\ and $\pvec_{\KL}$ from a fit with
the $B^0$ and \KL\ masses constrained to world-average values~\cite{PDG2006}.  
To make a match with the measured \KL\ direction
we construct the missing momentum $\pvec_{\rm
miss}$ from $\pvec_0$ and all charged tracks and neutral clusters other
than the \KL\ candidate.  We then project $\pvec_{\rm miss}$ onto $\pvec_{\KL}$, and
require the component perpendicular to the beam line, 
$p_{{\rm miss}\perp}^{\rm proj}$, to satisfy
$p_{{\rm miss}\perp}^{\rm proj}-p_{\KL\perp} > -0.5\ \gev$.  This value
was chosen to minimize the yield
uncertainty in the presence of background.

For \etapKzs\  we require $5.25<\mes<5.29\ \gev$ and $|\DE|<0.2$ \gev,
for \etapKzl\ we require  $-0.01<\DE<0.04$ \gev , and for all decays 
$|\dt|<20$ ${\rm ps}$, and, for the error on \dt, $\sigdt<2.5$ ${\rm ps}$. 

Background events arise primarily from random combinations of particles in
continuum $\epem\ra\qqbar$ events ($q=u,d,s,c$).  We reduce these with
requirements on the angle \thetaT\ between the thrust axis of the $B$
candidate in the \UfourS\ frame and that of the rest of the charged
tracks and neutral calorimeter clusters in the event.  
In the fit we discriminate further against \qqbar\ background with a
Fisher discriminant \xf\ that combines several variables that
characterize the production dynamics and energy flow in the event
\cite{AngMom}.  
For the \etaprg\ decays we require $|\cos\theta^{\rho}_{\rm dec}|< 0.9$
to reduce the combinatorial  background.
Here
$\theta_{\rm dec}^\rho$ is the angle between the momenta of the \rhoz\
daughter \pim\ and of the \etapr, measured in the \rhoz\ rest frame.

For \BetapKzl\ candidates we require that the cosine of the
polar angle of the total missing momentum in the laboratory
system be less than 0.95, to reject very forward \qqbar\ jets.
The purity of the \KL\ candidates reconstructed in the EMC is further
improved by a requirement on the output of a neural network (NN) that
takes cluster-shape variables as inputs. The NN was trained on MC
signal events and data events in the region \mbox{$0.02<\DE<0.04$
\gev}.  We check the performance of the NN on data  with \KL\ candidates in the
larger $\Bz \to J/\psi K^0_L$ data sample.

The average number of candidates found per selected event is between
1.08 and 1.32, depending on the final state. 
In the case of events with multiple candidates we choose the candidate with the smallest value of a $\chi^2$ 
constructed from the deviations from expected values of one or more of
the daughter resonance masses, or with the best decay vertex probability for the
$B$, depending on the decay channel.  Furthermore, in the \etapKzl\
sample, if several $B$
candidates have the same vertex probability, we choose the candidate with 
the \KL\ information taken  from, in order, EMC and IFR, EMC only, or IFR only.
From the simulation we find that this algorithm selects the correct-combination 
candidate in about two thirds of the events containing multiple candidates. 

We obtain the common \CP-violation parameters and signal yields for each
channel from a maximum likelihood fit with the input observables \DE,
\mes, \xf, and \deltat.  The selected sample sizes are given in the
first column of Table~\ref{tab:Results}.  
We estimate from the simulation a contribution to the  input
sample of less than 1.1 \% of
background from other charmless $B$ decay  modes. These events have final
states different from the signal, but similar kinematics, and exhibit 
broad peaks in the signal regions of some observables.  
We find that the \BB\ background component is needed only for the
channels with  \etaprg. 
We account for these with a separate component in the
probability density function (PDF).  For each component $j$ (signal,
\qqbar combinatorial background, or \BB\ background) and tagging
category $c$, we define a total probability density function for
event $i$ as:
\begin{equation}
{\cal P}_{j,c}^i \equiv {\cal P}_j ( \mes^i ) \cdot {\cal
P}_j ( \DE^i ) \cdot { \cal P}_j( \xf^i ) \cdot 
{ \cal  P}_j (\deltat^i, \sigma_{\deltat}^i;c)\,,
\end{equation}
except for \etapKzl\ for which ${\cal P}_j ( \mes^i )$ is omitted.
The factored form of the PDF is a good approximation
since linear correlations are small.

We write the extended likelihood function for all events of the  decay mode $d$ as
\begin{equation}
{{\cal L}_d} =  \prod_c  \exp{(-n_c)} \prod_i^{N_c} \left[ \sum_{j}n_j
f_{j,c}  {\cal P}_{j,c}^i  \right]\,,
\end{equation}
where $n_j$ is the yield of events of component $j$,  $f_{j,c}$ is the fraction of events
of component $j$ for each category $c$, 
$n_c =  n_{sig}f_{sig,c}+n_{q\bar{q}}f_{q\bar{q},c}+n_{B\bar{B}}f_{B\bar{B},c}$ 
is the number of events found by the fitter for category $c$, and $N_c$ is 
the number of events of category $c$ in the sample. When combining decay modes 
we form the grand likelihood ${\cal L}=\prod{\cal L}_d$. 
We fix both $f_{{\rm sig},c}$ and 
$f_{B\bar{B},c}$ to $f_{\bflav,c}$, the values measured with the large
sample of fully reconstructed $B^0$ decays into flavor eigenstates (\bflav\ sample)  \cite{Stefan}. 

The PDF ${ \cal P}_{\rm sig} (\dt,\, \sigdt; c)$, for each category
$c$, is the 
convolution of $F(\dt;\, c)$ (Eq.\ \ref{eq:FCPdef}) with the
signal resolution function (sum of three Gaussians) determined from the
\bflav\ sample.
The other PDF forms are: the sum of two Gaussians for ${\cal P}_{\rm
sig}(\mes)$ and ${\cal P}_{\rm sig}(\DE)$; the sum of three Gaussians for
${\cal P}_{\qqbar}(\dt; c)$ and ${\cal P}_{\BB}(\dt; c)$; an asymmetric  Gaussian with different
widths below and above the peak for ${\cal P}_j(\xf)$ (a small ``tail"
Gaussian is added for ${\cal P}_{\qqbar}(\xf)$); a linear
dependence for ${\cal P}_{\qqbar}(\DE)$ and a fourth-order
polynomial for ${\cal P}_{\BB}(\DE)$; for ${\cal
P}_{\qqbar}(\mes)$ and ${\cal
P}_{\BB}(\mes)$ the function
$x\sqrt{1-x^2}\exp{\left[-\xi(1-x^2)\right]}$,
with $x\equiv2\mes/\sqrt{s}$ and $\xi$ a free parameter \cite{Argus} and the same function plus a Gaussian, respectively.

For the signal and \BB\ background components we determine the PDF
parameters from simulation.  
We study large control samples of $B$ decays to
charm final states of similar topology to verify the simulated
resolutions in \DE\ and \mes, adjusting the PDFs to account for any
differences found.  
The \qqbar\ background parameters are free to vary in the final fit.
Thus, for the six
channels listed in Table \ref{tab:Results}, we perform a single fit with
93 free parameters: $S$, $C$, signal yields (6),
$\etapr_{\rho\gamma} K^0$ \BB\ background yields (2), continuum
background yields (6) and fractions (30), background \dt,
\mes, \DE, \xf\ PDF parameters (47). The parameters $\tau$ and 
$\deltamd$ are fixed to world-average values \cite{PDG2006}.  

We test and calibrate the fitting procedure by applying it to
ensembles of simulated experiments with \qqbar\ events drawn from the PDF into which
we have embedded the expected number of signal and \BB\ background
events randomly extracted from the fully simulated MC samples.  We find
negligible bias for $C$.  For $S$ we find and apply multiplicative
correction factors 
for bias from dilution due to cross-feed  from \BB\ background to
signal events  equal to 1.03 in 
the final states $\fetaprgKz_{\pi^+\pi^-}$, \fetapreppggkzl, and 
$\fetaprgKz_{\pi^{0}\pi^{0}}$.   

\begin{table}[!hb]
\caption{Results of the fits.  Subscripts for \etapr\ decay
modes denote \etapeppgg\ (1), \etaprg\ (2), and
\etapeppppp\ (3).
} 
\label{tab:Results}
\vspace*{-0.3cm}
\begin{center}
\begin{tabular}{lcccc}
\hline\hline
Mode                              &\# events &Signal yield  & $ S$       &         $C$      \\
\hline
$\etapr_1 K^0_{\pi^+\pi^-}$	&664		&$224\pm16$  &$\msp0.61\pm0.23$ &$-0.26\pm0.14$ \\
$\etapr_2 K^0_{\pi^+\pi^-}$	&11943	&$566\pm30$  &$\msp0.56\pm0.14$ &$-0.24\pm0.10$ \\
$\etapr_3 K^0_{\pi^+\pi^-}$	&177	&$73\pm9$    &$\msp0.89\pm0.35$ &$\msp0.14\pm0.25$ \\
$\etapr_1 K^0_{\pi^0\pi^0}$	&490	&$52\pm9$    &$\msp0.84\pm0.42$ &$-0.26\pm0.36$ \\
$\etapr_2 K^0_{\pi^0\pi^0}$	&13915	&$133\pm24$  &$\msp0.56\pm0.41$ &$\msp0.15\pm0.27$ \\
\hline
$\etapKzs$              &	&            &$\msp0.62\pm0.11$ &$-0.18\pm0.07$ \\
\hline
$\etapr_1\KL$		&4199	&$204\pm24$  &$\msp0.32\pm0.28$ &$\msp0.08\pm0.23$ \\
\hline
$\fetapKz$		&	&            &$\msp0.58\pm0.10$ &$-0.16\pm0.07$ \\
\hline\hline
\end{tabular}
\end{center}
\vspace*{-0.3cm}
\end{table}

\begin{figure}[!htb]
\hspace*{-0.5cm}
 \includegraphics[angle=0,width=0.5\textwidth]{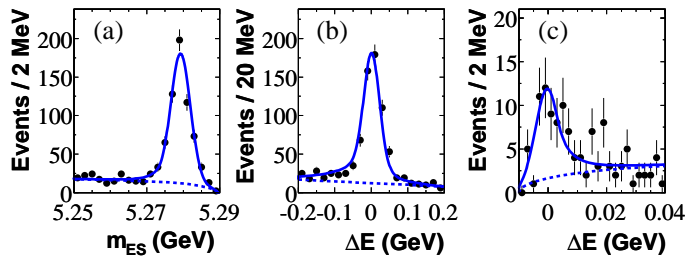}\\
 \caption{\label{fig:projMbDE}
Distributions projected (see text)
onto (a) \mb\ and (b) \DE\ for \etapKzs\
candidates, and (c) \DE\ for \etapKzl\ candidates.  The solid lines shows
the full fit result and the dashed lines show the background contributions.}
\end{figure}

\begin{figure}[!htb]
  \begin{center}
   \includegraphics[width=0.23\textwidth]{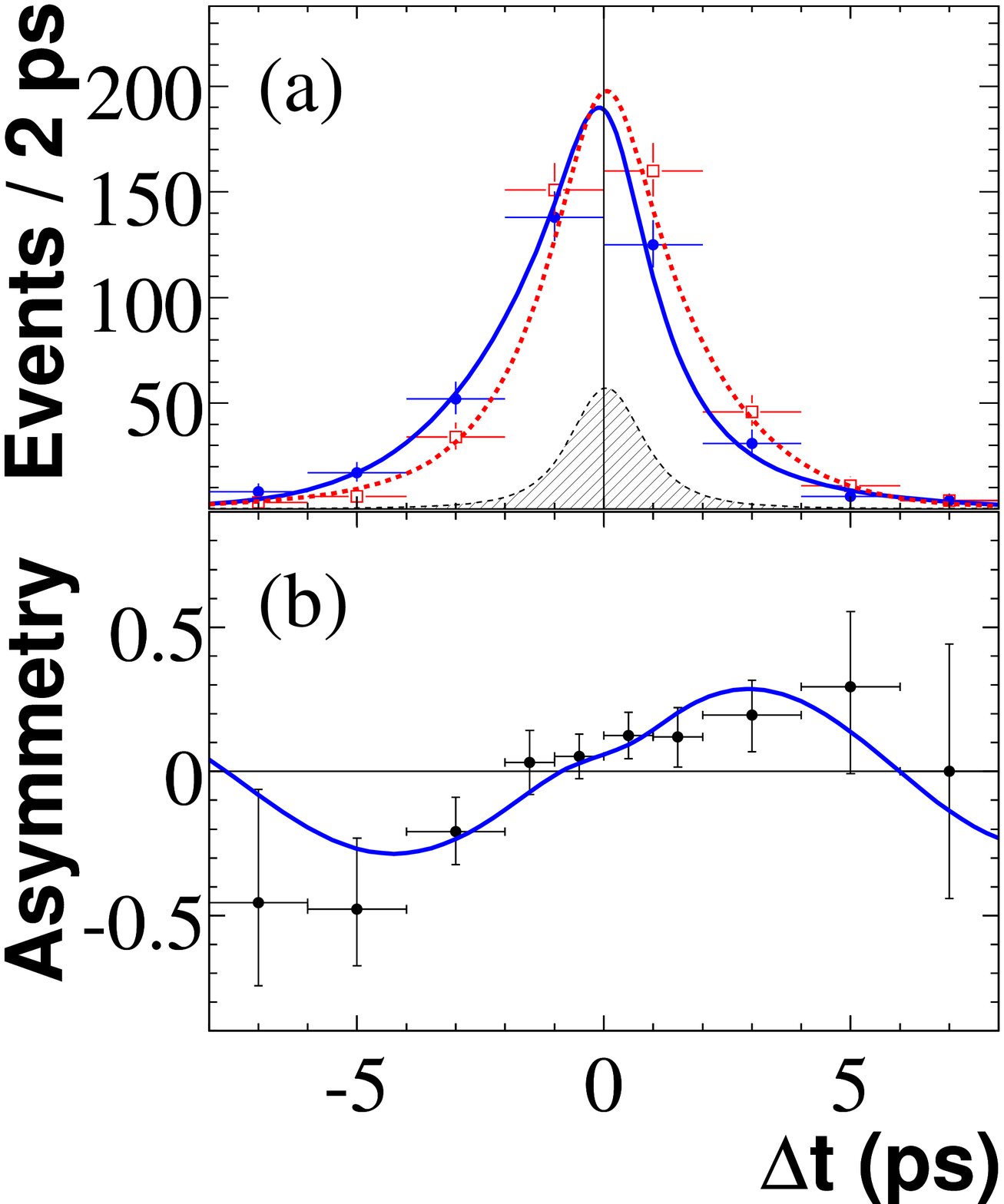}
   \includegraphics[width=0.23\textwidth]{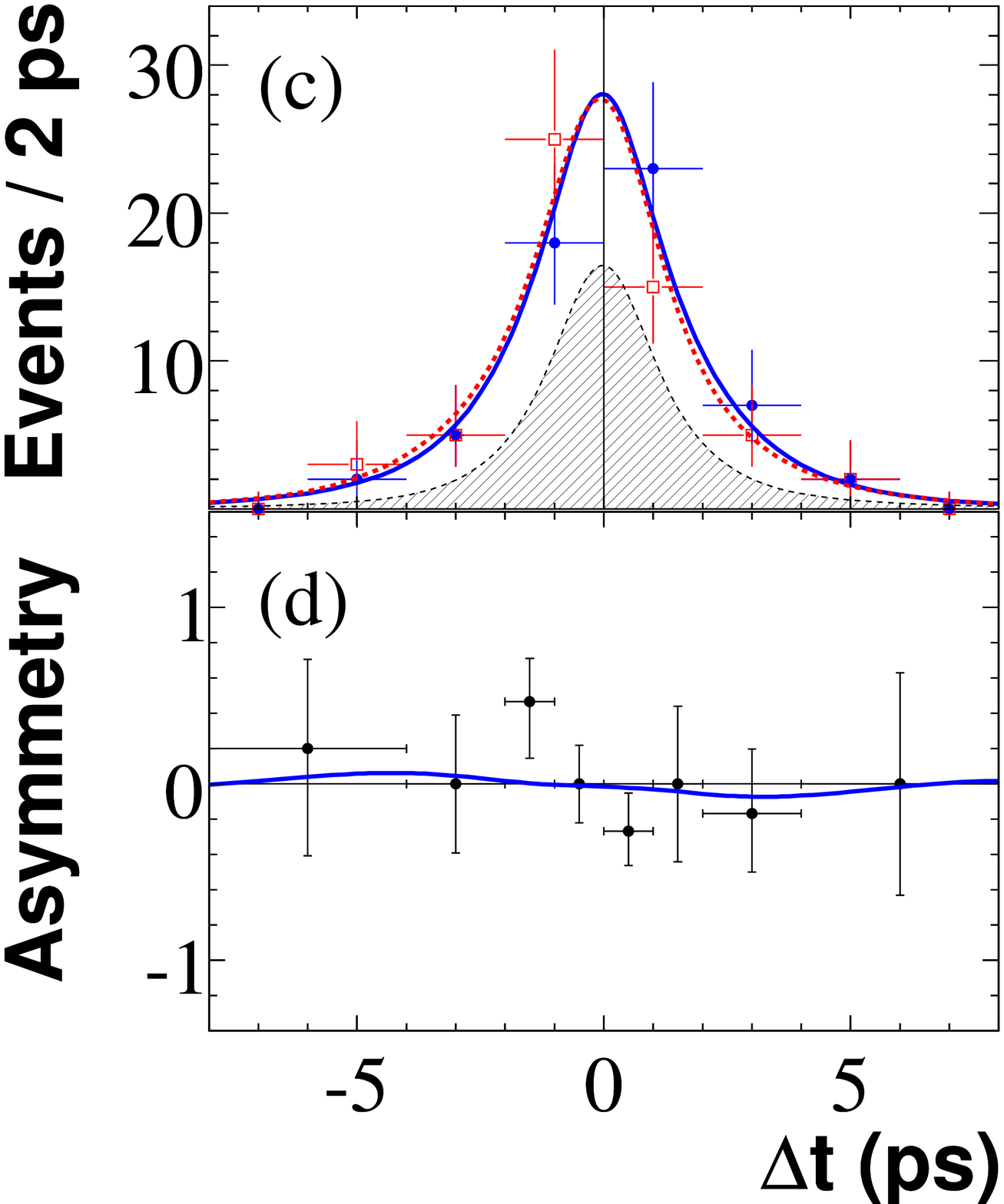}
\end{center}
  \vspace*{-0.5cm}
 \caption{Projections (see text)
onto $\Delta t$ for (a) \etapKzs\ and (c) \etapKzl\ of the data (points 
with  error bars for \Bz\ tags ($N_{\Bz}$) in red empty rectangles 
and \Bzb\ tags ($N_{\Bzb}$) in in blue solid circles), fit function (red dashed and
blue solid lines for \Bz\ and \Bzb\ tagged events, respectively), and background
function (black shaded regions). We show the raw asymmetry,
$(N_{\Bz}-N_{\Bzb})/(N_{\Bz}+N_{\Bzb})$,
for (b) \etapKzs\ and (d) \etapKzl; the  lines represent the fit
functions.}
  \label{fig:DeltaTProj}
\end{figure}

Results from the fit for the signal yields and the \CP parameters $S$
and $C$ are presented in Table \ref{tab:Results}.
In Fig.\ \ref{fig:projMbDE}\ we show the projections onto
\mb\ and \DE\ for 
a subset of the data for which the ratio between 
the likelihood of  signal events   and the sum of likelihoods of
signal and  background events
(computed without the variable plotted) exceeds a mode-dependent
threshold that optimizes the sensitivity.
In Fig.~\ref{fig:DeltaTProj} we give the $\Delta t$ and asymmetry
projections of the 
events selected as for Fig.~\ref{fig:projMbDE}.
We measure a correlation of $3.2\%$ between $S$ and $C$ in the fit.

We perform several crosschecks of our analysis technique including
 time-dependent  fits for $B^+$
decays to the charged final states $\fetapreppggk$, $\fetaprgKp$, 
and $\fetapreppthrpik$; fits removing one fit variable
at a time; fits without \BB\ PDFs; fits with multiple \BB\ components; 
fits allowing for non-zero \CP information in \BB events; fits with $C
= 0$. In all cases, we find results consistent with expectation.  The
value $S=0.62\pm0.11$ for \etapKzs\ differs from our previous measurement
$S=0.30\pm0.14$ \cite{Previous} due to the improved event
 reconstruction
(with a contribution of +0.08)
and selection (+0.12) and to the additional data collected (+0.12). With
 a model of the data sample changes  introduced by our revised event reconstruction
 and new data, we find that our current result has a statistical
 probability of 35\% (50\%) for an assumed true value of $S$ of 0.61 (0.70).

%  The former contributes
%a change of 0.20, mostly due to events added or removed from the
%original dataset, which, based on simulation, is consistent with expectations 
%for statistical fluctuations.  The new data contributes the rest of the 
%difference.

We have studied the
systematic uncertainties arising  from several sources (in decreasing order of
magnitude): variation of the signal
PDF shape parameters within their errors, modeling of the signal \dt\
distribution, use of \dt\ signal parameters
from the \bflav\ sample, interference between the CKM-suppressed
$\bar{b}\to\bar{u} c\bar{d}$ amplitude and the favored $b\to c\bar{u}d$
amplitude for some tag-side $B$ decays
\cite{dcsd}, \BB\ background, SVT alignment, and position and
size of the beam spot. The \bflav\ sample is used  to determine the errors
associated with the signal \dt\ resolutions, tagging efficiencies, and
mistag rates.  We take the uncertainties in $\tau_B$ and \deltamd\ from 
published measurements \cite{PDG2006}. 
Summing all systematic errors in quadrature, we obtain 0.03 
for $S$ and 0.03 for $C$.

In conclusion, we have used a sample containing  $1252\pm50$ flavor-tagged \etapKz\ 
events to measure the time-dependent \CP violation parameters,
 $\skz = \SetapKz$ and $\ckz = \CetapKz$.  
This sample is 2.1 times as large as that of our previous measurement 
\cite{Previous}.
Our result for
\skz\ is consistent with the world average of \stwob\ measured in $\Bz\ra
J/\psi\KS$ \cite{PDG2006}. We observe mixing-induced \CP\ violation  in $B^0$ decays to \etapKz\ 
with a significance (systematic uncertainties included) of 5.5
standard deviations. 
Our result for 
direct-\CP violation is 2.1 standard deviations from zero. 

\input acknow_PRL.tex

\end{document}

%% file: authors_sep2006.tex
%% author list as of 01-Sep-2006 (603 authors)
%
\author{B.~Aubert}
\author{M.~Bona}
\author{D.~Boutigny}
\author{F.~Couderc}
\author{Y.~Karyotakis}
\author{J.~P.~Lees}
\author{V.~Poireau}
\author{V.~Tisserand}
\author{A.~Zghiche}
\affiliation{Laboratoire de Physique des Particules, IN2P3/CNRS et Universit\'e de Savoie, F-74941 Annecy-Le-Vieux, France }
\author{E.~Grauges}
\affiliation{Universitat de Barcelona, Facultat de Fisica, Departament ECM, E-08028 Barcelona, Spain }
% ...
\collaboration{The \babar\ Collaboration}
\noaffiliation

%% file: acknow_PRL.tex
We are grateful for the excellent luminosity and machine conditions
provided by our \pep2\ colleagues, 
and for the substantial dedicated effort from
the computing organizations that support \babar.
The collaborating institutions wish to thank 
SLAC for its support and kind hospitality. 
This work is supported by
DOE
and NSF (USA),
NSERC (Canada),
IHEP (China),
CEA and
CNRS-IN2P3
(France),
BMBF and DFG
(Germany),
INFN (Italy),
FOM (The Netherlands),
NFR (Norway),
MIST (Russia),
MEC (Spain), and
PPARC (United Kingdom). 
Individuals have received support from the
Marie Curie EIF (European Union) and
the A.~P.~Sloan Foundation.